\documentclass[%
 reprint,
 amsmath,amssymb,
 aps,
prb,
]{revtex4-2}

\usepackage{graphicx}
\usepackage{dcolumn}
\usepackage{bm}
\usepackage[colorlinks=true, allcolors=blue]{hyperref}

\usepackage{multirow}
\usepackage{makecell}
\usepackage{textcomp}
\usepackage[T1]{fontenc}
\usepackage{mathptmx}
\usepackage{fancyhdr}

\fancypagestyle{plain}{
    \fancyhf{}
    \fancyhead[c]{Frontiers of Physics\\https://doi.org/10.1007/s11467-021-1077-6}
    \fancyhead[RO,RE]{Front. Phys.\\16(4), 43505 (2021)}
    \fancyfoot[LE,LO]{~\thepage~}
}
\newcommand\po{$\bm{\pi}$}
\newcommand\so{$\bm{\sigma}$}

\newcommand\ti{\textit{i.e.}}
\newcommand\fe{\textit{e.g.}}
\newcommand\et{\textit{et al.}}

\begin{document}

\preprint{APS/123-QED}

\title{\textit{Ab initio} study of anisotropic mechanical and electronic properties of strained carbon-nitride nanosheet with interlayer bonding}

\author{Hao Cheng}
 \affiliation{Department of Physics and the Collaborative Innovation Center for Optoelectronic Semiconductors and Efficient Devices, Xiamen University, Xiamen 361005, China}
 \affiliation{Department of Physics, Xiamen University Malaysia, 439000, Sepang, Selangor, Malaysia}
\author{Jin-Cheng Zheng}
 \email[E-mail:]{jczheng@xmu.edu.cn}
 \affiliation{Department of Physics and the Collaborative Innovation Center for Optoelectronic Semiconductors and Efficient Devices, Xiamen University, Xiamen 361005, China}
 \affiliation{Department of Physics, Xiamen University Malaysia, 439000, Sepang, Selangor, Malaysia}

\date{March 27, 2021}

\begin{abstract}
Due to the noticeable structural similarity and being neighborhood in periodic table of group-IV and -V elemental monolayers, 
whether the combination of group-IV and -V elements could have stable nanosheet structures with optimistic properties has attracted great research interest.
In this work, we performed first-principles simulations to investigate the elastic, vibrational and electronic properties of the carbon nitride (CN) nanosheet in the puckered honeycomb structure with covalent interlayer bonding.
It has been demonstrated that the structural stability of CN nanosheet is essentially maintained by the strong interlayer \so\ bonding between adjacent carbon atoms in the opposite atomic layers. 
A negative Poisson's ratio in the out-of-plane direction under biaxial deformation, and the extreme in-plane stiffness of CN nanosheet, only slightly inferior to the monolayer graphene, are revealed.
Moreover, the highly anisotropic mechanical and electronic response of CN nanosheet to tensile strain have been explored.
\begin{description}
\item[DOI]
\href{https://doi.org/10.1007/s11467-021-1077-6}{https://doi.org/10.1007/s11467-021-1077-6}
\end{description}
\end{abstract}

\maketitle


\section{Introduction}
The fascinating mechanical and electronic properties of monolayer graphene and its heterostructures,
derived from its unique hexagonal symmetry with $sp^2$ intralayer bonding, 
such as high carrier mobility, massless Dirac fermions, 
and ultrastrength in-plane stiffness, 
have attracted significant research interest in two-dimensional (2D) materials for the last decade \cite{RN888,RN889,RN890}.
Not only have the extraordinary properties of monolayer graphene been revealed, 
but also the existence of other group-IV elements forming honeycomb lattices, \ti, silicene and germanene, has been demonstrated \cite{RN891,RN892}.
However, the electronic application of group-IV element monolayer is limited by the band gap closure around the Fermi level \cite{RN893}.
Recent synthesis of phosphorene has demonstrated the actual existence of group-V elemental monolayer and its intriguing properties, 
such as direct band gap, high carrier mobility, and prevalent optical properties have been revealed \cite{RN894, RN895,RN932}.
Besides, the theoretical predictions of stability, electronic structures \cite{zheng2002structural, RN457} and experimental synthesis \cite{RN457} of group-III element monolayers have been systematically studied.
Because of their noticeable structural similarity and being neighborhood in periodic table of group-IV and -V elemental monolayers, 
one would ask, whether the compounds consisting of group-IV and -V elements could have stable nanosheet structures with optimistic properties.
Plentiful attempts have been directed towards synthesizing group IV-V compounds, \fe, carbon nitride (CN) and carbon phosphide (CP), 
from three-dimensional (3D) bulk phases to 2D nanosheets \cite{RN933}.
Some experiment results have manifested the existence of $\beta$-$\rm{C_3N_4}$ showing comparable hardness to diamond \cite{RN896, RN897, RN898}. 
Moreover, the possibility of a wide range of other nitrogen concentration is shown in other experimental data \cite{RN899, RN900}.
The synthesis of amorphous CP films using radio frequency plasma deposition method has been reported by Pearce \et\ \cite{RN901} 
Different C/P ratios can be achieved by adjusting the ratio of PH3/CH4 gas during the synthesis.
In other experiments, the pulsed laser deposition \cite{RN902, RN903} and magnetron sputtering techniques \cite{RN904} were used in producing CP films.

Several possible crystal structures for CN bulk phase have been suggested via theoretical study, including the eight structures considered by Cote and Cohen,
among which the GaSe layered phase (honeycomb crystal) is assumed to be the most energetically favorable structure \cite{RN905}. 
Besides that, in our earlier work \cite{RN906}, it is predicted that the GaSe layered phase are energetically favorable for all group IV-V compounds except for SnSb, 
and it shows semiconductor characters while other structures show metallic properties.
For 2D nanosheets, Wang \et\ suggested three kinds of potential structures for CP which they calculated super carrier mobility and strong anisotropy \cite{RN907}.
On the other hand, the electronic and thermal properties of multi-layer nanosheets and its heterostructures are substantially affected by the interlayer interaction \cite{RN908, RN909, RN934, RN910, jiajing2021}.
There are generally two ways for layer interactions, Van der Walls (vdW) force and covalent bonding.
The covalent bonding naturally brings in stronger cross-plane coupling than the weak vdW force.
Recent experiments have shown that the \so\ bonding of C atoms between the layers of graphite could be induced by visible-light irradiation \cite{RN911}. 
The cooperative and nonlinear formation mechanisms of this excited interlayer \so\ bonds are demonstrated \cite{RN912}. 
The influence of these interlayer interaction to the thermal properties of multi-layer graphene has been conducted both experimentally and theoretically \cite{RN913, RN914, RN915}.

In this paper, we performed first-principles simulations to investigate the elastic, vibrational and electronic properties of the CN nanosheet in the puckered honeycomb structure with covalent interlayer bonding. 
It has been demonstrated that the structure stability of CN nanosheets is essentially maintained by the strong interlayer \so\ bonding between two adjacent carbon atoms in the opposite atomic layers.
Our numerical results show a negative Poisson's ratio in the out-of-plane direction under biaxial deformation.
We attribute this abnormal Poisson's ratio to the introduction of additional degrees of out-of-plane atomic relaxations by the formation of covalent interlayer bonding in CN nanosheet.
With the application of various uniaxial and biaxial tension strains, 
the highly anisotropic mechanical and electronic response of CN nanosheet to the the strain has been explored.
It is identified that the tensile strength of CN nanosheets are dictated by the instability of out-of-plane transverse phonon branch ZA for armchair tension $\epsilon_{xx}$, but in-plane transverse phonon branch TA for the zigzag tension $\epsilon_{yy}$. 
Our calculations indicate that the indirect band gap of CN nanosheet could be significantly tailored through strain engineering which manifests its potential application in the optoelectronic fields.

\section{Method}
\label{method}
\begin{figure}[t]
\includegraphics[width=\columnwidth]{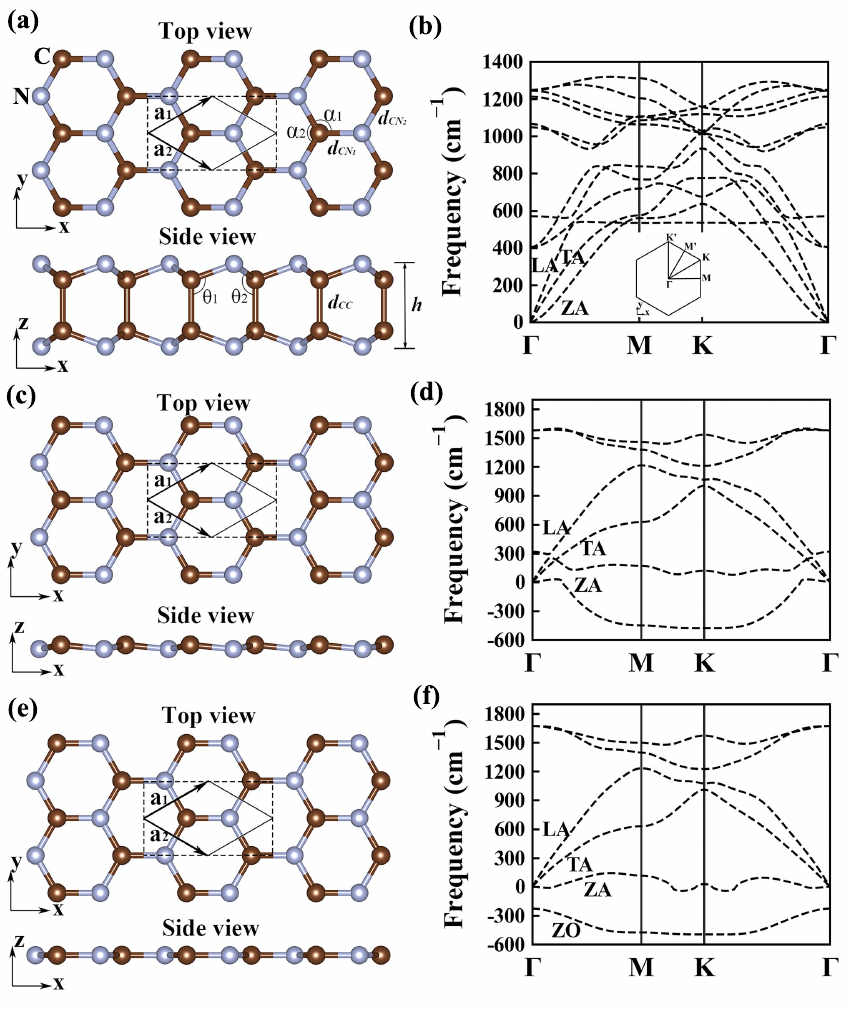}
\caption{
\label{structure}
The crystal structures including top and side views of CN nanosheets forming (a) LB geometry with covalent interlayer bonding $d_{CC}$, (c) LB geometry without interlayer bonding (monolayer), (e) PL geometry (monolayer). The dark brown and light blue spheres are C and N atoms, respectively. 
The Bravais lattice vectors of primitive are indicated by $\bm{a_1}$, $\bm{a_2}$ and vector directions are represented with arrows.
The x and y directions correspond to the armchair and zigzag directions, respectively.
An eight-atom unit cell for uniaxial tension calculation is denoted with dashed box.
The corresponding ground state phonon dispersion curves of these three atomic configurations are presented in (b), (d), and (f). The acoustic phonon and imaginary phonon are indicated. The unstable phonon branch is ZA for (c), ZO and ZA for (e). Inset shows the deformed first BZ under uniaxial tension.
}
\end{figure}

We carried out the first-principles calculations based on the density functional theory (DFT) as implemented in the Quantum Espresso simulation package \cite{RN922}.
The Perdew-Burke-Ernzerhof (PBE) exchange-correlation functional \cite{RN916} along with the projector augmented wave (PAW) pseudopotentials \cite{RN917} were adopted for the self-consistent total energy and electronic-structure calculations.
The kinetic energy cutoff for wave functions and charge density were set to be 85\,Ry and 595\,Ry respectively.
To model CN nanosheets, a four-atom primitive cell containing two carbon atoms and two nitrogen atoms was chosen with periodic boundary condition applied (see Fig.~\ref{structure}).
Specifically, for stress-strain calculations, 
an eight-atom orthogonal unit cell was employed to achieve uniaxial stress along the armchair and zigzag direction (Fig.~\ref{structure}, dashed boxes).
A Monkhorst-Pack grid of ($25\times25\times1$) $k$-points was used to sample the first reciprocal Brillouin Zone (BZ).
The energy convergence criteria for electronic and ionic iterations were set to be $\mathrm{10^{-10}}$\,Ry and $\mathrm{10^{-6}}$\,Ry, respectively.
All atomic positions and unit cell were optimized until the atomic forces were less than $\mathrm{10^{-5}}\,\mathrm{Ry/au}$. 
A minimum of 20\,\AA\ vacuum spacing in the out-of-plane direction was applied to prevent interaction from adjacent supercells.
The phonon dispersion curves were calculated by diagonalizing the dynamical matrix based on the density functional perturbation theory.

It is well known that the PBE functional may underestimate the calculated band gaps.
In this work, the frequency-dependent $GW$ quasiparticle calculations were performed to correct the ground state band gaps obtained by PBE. 
Furthermore, a scissors operator method has been implemented to correct band gaps under tensile strains \cite{RN985,RN986}.
The $GW$ calculations were conducted as a ‘‘one-shot’’ correction to self-consistent PBE calculations. 
This approach, denoted as $G_0W_0$, has yielded remarkably accurate band structures for many materials \cite{RN956,RN954,RN955}.
Various tests were carried out to achieve convergence regarding $k$-points, vacuum level, number of bands,  and kinetic energy cutoff. 
Final $G_0W_0$ corrections on the PBE bands gaps were obtained by using ($17\times17\times1$) $k$-points in BZ, 25\,\AA\ vacuum spacing, 120 bands and default cut-off potential for $G_0W_0$.

In contrast to the flatness of graphene, the CN nanosheet we proposed here is a puckered honeycomb structure with covalent interlayer bonding.
The bravais lattice vectors of primitive cell are defined as $\bm{a_1}$ and $\bm{a_2}$, shown in Fig.~\ref{structure}(a).
The armchair and zigzag directions are along the $x$ and $y$ directions, respectively.
Under uniaxial tension, the original lattice symmetry is broken, thus the structure parameters along armchair direction and zigzag direction are different (the subscripts are assigned to distinguish them).
The intralayer bonding $d_{CN_1}$ and $d_{CN_2}$ are formed between carbon and nitrogen atoms in the same atomic layer, while the interlayer bonding $d_{CC}$ is formed between two adjacent carbon atoms in the opposite atomic layers.
 The in-plane N-C-N bond angles is named as $\alpha_1$ and $\alpha_2$, and out-of-plane N-C-C bond angle is named as $\theta_1$ and $\theta_2$, respectively.
The layer height of CN nanosheet is characterized by the atomic distance between two nitrogen atoms sited opposite in the out-of-plane direction, \ti, $h$.
The first BZ is deformed under the uniaxial tension with high symmetry points splitting up, as shown in Fig.~\ref{structure}(b).

For the stress-strain relations, in order to make a direct comparison of 2D nanosheets with experiments and other theoretical results, 
we need to rescale the supercell stress by $\eta=Z/d_0$ to obtain the equivalent stress, 
where $Z$ is the cell length in the direction perpendicular to atomic layers and $d_0$ is the effective thickness of nanosheet. 
In this work, we take $d_0=5.69$\,\AA\ for CN nanosheet which is the equilibrium interlayer spacing of CN bulk phase calculated in previous work by Cohen \et\ \cite{RN905}
It is worthwhile to mention that the vdW interactions between adjacent atomic layers of CN nanosheet are neglected in our calculations because those vdW corrections are assumed to be much weaker than the covalent interlayer bonding $d_{CC}$ of CN nanosheet \cite{RN908}.

\begin{figure}[t]
\includegraphics[width=\columnwidth]{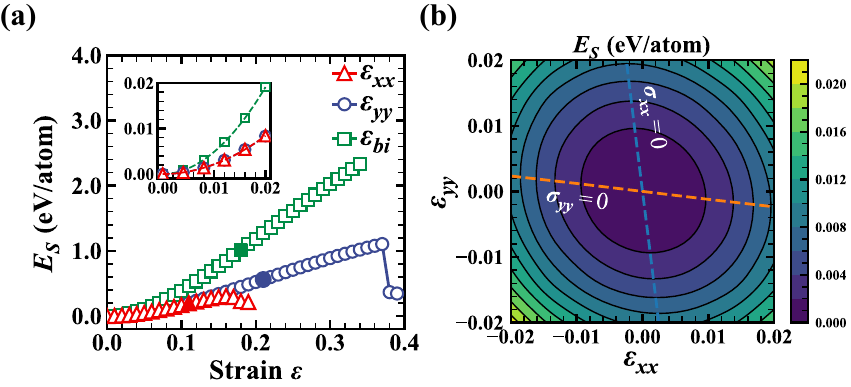}
\caption{
\label{strain energy}
(a) The strain energy $E_S$ of CN nanosheet as functions of tensile strain $\epsilon_{xx}$, $\epsilon_{yy}$, and $\epsilon_{bi}$ in uniform deformation region. The solid triangles, circles and squares indicate the condition where the peak stress could be attained along armchair, zigzag and biaxial tensions, respectively. Inset shows the polynomial regression of the initial strain curves for obtaining the corresponding elastic moduli.
(b) The contour plot of strain energy $E_S$ with respect to the ($\epsilon_{xx}$, $\epsilon_{yy}$) in the harmonic region. 
The energy-strain relation is fitted as quadratic polynomial $E_S=a_1\epsilon_{xx}^2+b_1\epsilon_{yy}^2+c_1\epsilon_{xx}\epsilon_{yy}$ with $a_1=b_1=14360.1$\,eV, and $c_1=3583.7$\,eV. 
The dashed lines denoted the uniaxial deformations along x and y directions, respectively.
}
\end{figure}

\section{Results}
We first optimize the equilibrium structural parameters of CN nanosheets in 2D honeycomb lattices. 
In present work, we consider three atomic configurations in accordance with the rate of buckling in the out-of-plane direction, \ti, low-buckled (LB) (or puckered) geometry with covalent interlayer bonding, LB geometry without interlayer bonding, and planar (PL) geometry. 
The structural stabilities of these three atomic configurations are examined by calculating the phonon dispersion curves.

Figure.~\ref{structure} shows the optimized structure models along with the ground state phonon dispersion curves of these three atomic configurations of CN nanosheets.
While the LB geometry without interlayer bonding and PL geometry both show imaginary (negative) frequencies in the phonon dispersion, 
the structural stability is exclusively attained by the LB geometry with covalent interlayer bonding $d_{CC}$.
Compared to the planar geometry, the out-of-plane transverse phonon branch ZO is stabilized through the buckling of C-N atomic layers in the LB geometry [Fig.~\ref{structure}(d) and \ref{structure}(f)].
The planar $sp^2$ orbital is slightly dehybridized with perpendicular $p_z$ orbital to form a $sp^3$-like orbital as a result of the buckling. 
This buckled stabilization mechanism could be interpreted as the in-plane \po\ bonding formed between C and N atoms is not strong enough to maintain the planar geometry by comparison with the C-C bonding in the monolayer graphene \cite{RN925}.
However, it is still unstable with respect to the out-of-plane transverse phonon branch ZA in the LB geometry without interlayer bonding.
Through the coupling of two symmetrically buckled atomic layers in hexagonal lattice, 
as the formation of covalent interlayer \so\ bonds between two adjacent C atoms, 
the CN nanosheets is stabilized with fourfold bonding of C atoms and threefold bonding of N atoms [Fig.~\ref{structure}(b) and \ref{structure}(d)].
The coordination number of C and N atoms are consistent with the number of valence electrons needed to fully fill their electronic shell respectively. 
In contrast to the weak vdW interactions between few layers planar graphene, 
this strong interlayer \so\ bonding between C atoms together with the buckled C-N atomic layers maintain the structure stability of CN nanosheet.
Besides that, It is further confirmed by the comparison of the calculated cohesive energy among these three atomic configurations. 
The cohesive energy is defined as $E_{coh} = [E_t(C)+E_t(N)-E_t(CN)]/2$, 
where $E_t(CN)$ is the total energy per pair atoms of optimized CN nanosheet; 
$E_t(C)$ and $E_t(N)$ are the total energies of isolated C and N atoms corresponding to nonmagnetic state.
The cohesive energy of LB geometry with covalent interlayer bonding, $E_{coh}=8.26$\,eV/atom, turns out to be much larger than other two atomic configurations without interlayer bonding (Table SI, supplementary material).
Consequently, from the indication of structure stability analysis above, our theoretical results below will be built based on the LB geometry with covalent interlayer bonding $d_{CC}$.

\begin{table}[t]
\caption{
\label{structure parameters}
The calculated structure parameters of CN nanosheet.
The value of angle between intralayer bond C-N-C and interlayer bond N-C-C, $\alpha$ and $\theta$; 
C-N and C-C bond length, $d_{CN}$ and $d_{CC}$; 
lattice constant, $a$; 
layer height, $h$;
cohesive energy, $E_{coh}$; 
are given.}
\begin{ruledtabular}
\begin{tabular}{ccccccccc}
Structure & 
\makecell{$\theta$\\(\textdegree)} &
\makecell{$\alpha$\\(\textdegree)} & 
\makecell{$d_{\mathrm{CN}}$\\(\AA)} & 
\makecell{$d_{\mathrm{CC}}$\\(\AA)} & 
\makecell{$a$\\(\AA)} &
\makecell{$h$\\(\AA)} &
\makecell{$E_{coh}$\\(eV/atom)} & 
\\\hline
CN & 110.3 & 108.6 & 1.46 & 1.63 & 2.38 & 2.65 & 8.36\\
\end{tabular}
\end{ruledtabular}
\end{table}

Table.~\ref{structure parameters} summarizes the calculated equilibrium structural parameters of CN nanosheet.
It is interesting to compare the lattice parameters of CN nanosheet with those of graphene and other hexagonal lattice structures.
The bond length of intralayer C-N bond of CN nanosheet, $d_{CN}=1.46$\,\AA, 
is approaching that of the C-C bond in monolayer graphene (1.44\,\AA) and B-N bond in boron nitride (1.45\,\AA), both having very high in-plane strength, 
but much shorter than the Mo-S bond in $\mathrm{MoS_2}$\ (2.42\,\AA) and all other potential hexagonal structure of binary monolayer of group-IV elements and group III-V compounds reported by \c{S}ahain \et\cite{RN925}\@
As for the interlayer C-C bond, $d_{CC}=1.63$\,\AA, 
it is a bit longer than the bond length of $sp^3$ hybridized orbital in diamond (1.53\,\AA). 
In addition, the in-plane bond angle $\alpha$ and out-of-plane bond angle $\theta$ are 108.6\textdegree\ and 110.3\textdegree\ respectively, 
which are quite close to that of 109.5\textdegree\ in diamond. 
Briefly, the strong interlayer \so\ bonding together with the hybridized $sp^3$-like orbital underlie unusual in-plane and out-of-plane mechanical strength of CN nanosheet compared to other traditional 2D materials.

\begin{figure}[t]
\includegraphics[width=\columnwidth]{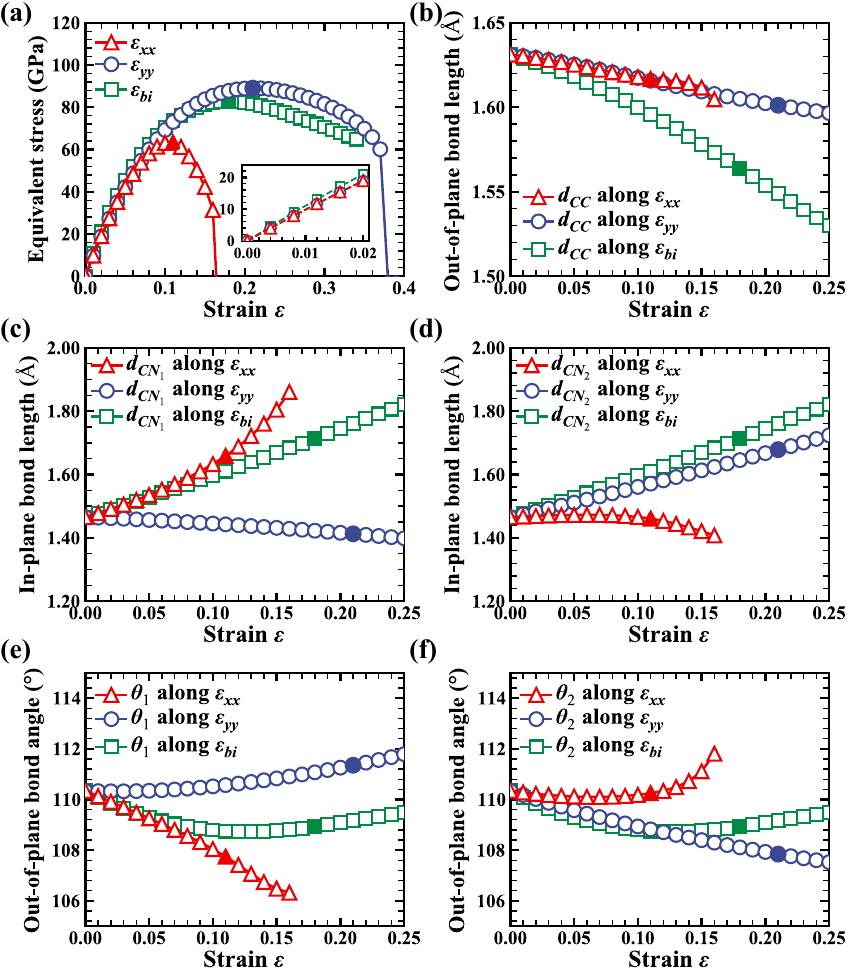}
\caption{
\label{strain stress}
(a) Calculated equivalent stress ($d_0$=5.69\AA) versus tensile strain $\epsilon_{xx}$, $\epsilon_{yy}$ and $\epsilon_{bi}$.  
(b) The variation of the out-of-plane bond length $d_{CC}$ with respect to the applied tensile strain.
The in-plane bond length (c) $d_{CN_1}$ and (d) $d_{CN_2}$  as functions of the applied tensile strain.
The out-of-plane bond angle (e) $\theta_1$ and (f) $\theta_2$ as functions of the applied tensile strain.
The solid triangles, circles and squares indicate the condition where the peak stress could be attained along armchair, zigzag and biaxial tensions, respectively.
} 
\end{figure}

\begin{figure}[b]
\includegraphics[width=2.5in]{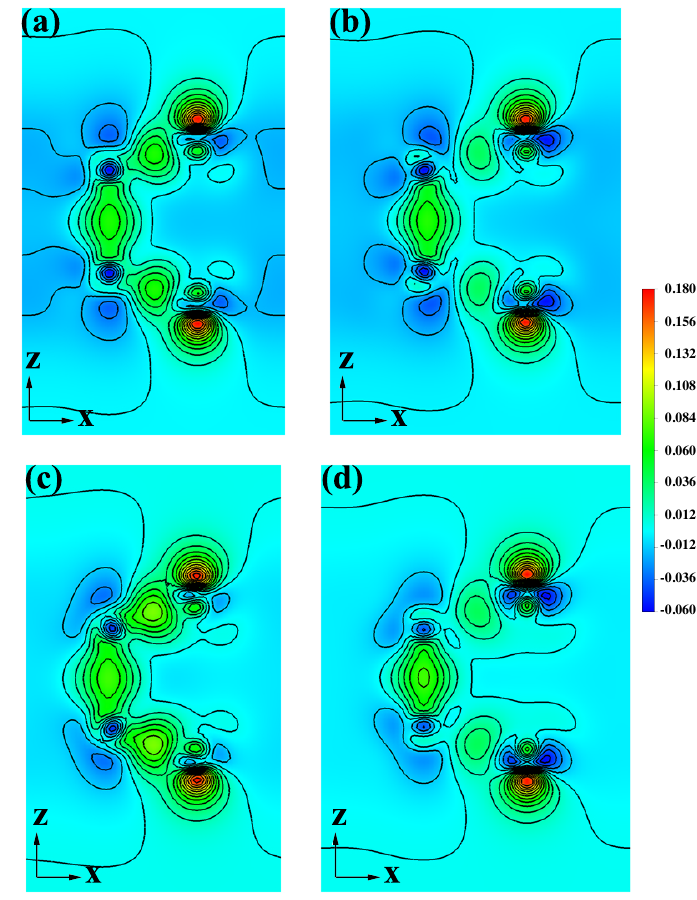}
\caption{
\label{charge density}
Side view of the deformation charge density of CN nanosheet at (a) equilibrium state, (b) uniaxial tension $\epsilon_{xx}=0.11$, (c) uniaxial tension $\epsilon_{yy}=0.21$, and (d) biaxial tension $\epsilon_{bi}=0.18$ where peak stress could be obtained.
The blue and red colored isosurfaces represent the depletion and accumulation of charge, respectively.
The contour interval is set to be 0.015 electron/$\mathrm{Bohr^3}$.
} 
\end{figure}

To further explore the impact of strain on vibrational, mechanical and electronic properties of CN nanosheet, 
we then stretch the eight-atom unit cell of CN with a series of incremental tensile strains including uniaxial tension along $x$ (armchair) direction $\epsilon_{xx}$, 
uniaxial tension along $y$ (zigzag) direction $\epsilon_{yy}$, and biaxial tension $\epsilon_{bi}$ ($\epsilon_{xx}=\epsilon_{yy}$). 
The atomic positions for each step are taken from the relaxed coordinates of the former step to ensure the continuity of strain path. 
For the uniaxial tension, the optimized structures are attained by simultaneously relaxing the strain components orthogonal to the applied strain. 
The variation in strain energies, $E_{S}$, 
by subtracting the total energy per atom of the deformed structure from the equilibrium one, 
is obtained as functions of elastic strain in both harmonic and anharmonic domain.
Owing to the hexagonal symmetry of honeycomb structure, 
CN nanosheet shows isotropic strain energy response at small strains [inset in Fig.~\ref{strain energy}(a)].
In Fig.~\ref{strain energy}(b), we show the contour plot of strain energy as functions of strains ($\epsilon_{xx}, \epsilon_{yy}$). 
The energy-strain relation is then obtained as a quadratic polynomial  $E_S=a_1\epsilon_{xx}^2+b_1\epsilon_{yy}^2+c_1\epsilon_{xx}\epsilon_{yy}$, 
where $a_1$, $b_1$, and $c_1$ are parameters to be fitted. 
With the small strain limit, we set $a_1=b_1$ as CN nanosheet is elastically isotropic within the plane.
The elastic stiffness constants can then be expressed in terms of these fitting parameters, \ti, 
$C_{11}=2a_1/(h \cdot A_0)$; $C_{12} = c_1/(h \cdot A_0)$.
Hence, the effective Young's modulus can be obtained as, $E=C_{11}\cdot[1-(C_{12}/C_{11})^2]=[2a_1-c_1^2/2a]/(h\cdot A_0)$, 
where $A_0$ are the equilibrium surface area of the system.
The CN nanosheet is calculated to have an effective Young's modulus $E=1013$\,GPa which is comparable to that of 1\,TPa in monolayer graphene \cite{RN930,RN957}. 

Furthermore, we calculate the ideal strength of CN nanosheet under tensile strain.
Figure.~\ref{strain stress}(a) shows the equivalent stress as functions of tensile strains for CN nanosheet.
Similar to the strain energy response, CN nanosheet has isotropic in-plane elastic response at small strains. 
There is no observable difference of the elastic response between the uniaxial tension $\epsilon_{xx}$ and $\epsilon_{yy}$ when strains are less than 0.02. 
While, at large strains, due to the broken of lattice symmetry, the elastic response in armchair direction becomes distinct from zigzag direction. 
The maximum stress for uniaxial tension along armchair direction is $\sigma_{xx}=63\,\mathrm{GPa}$, at $\epsilon_{xx}=0.11$.
Compared to that, CN nanosheet is stronger in the zigzag direction with maximum stress  $\sigma_{yy} = 89\,\mathrm{GPa}$, at $\epsilon_{yy}=0.21$.
Thus, we predict that the zigzag direction is 41\% stronger than the armchair direction and can sustain 91\% more strain.
For the biaxial tension, the maximum stress is 82\,GPa, at $\epsilon_{bi}=0.18$, which is between that along the armchair and zigzag tensions.
It is worth mentioned that the ideal strength of CN nanosheet in armchair and zigzag direction are both smaller than those of 110\,GPa and 121\,GPa  in monolayer graphene \cite{RN160}, 
but approaching that of 88\,GPa and 102\,GPa in single-layer hexagonal BN \cite{RN928}.
Moreover, we have explored the responses of the atomic relaxations, \ti, bond length and bond angle, 
with respect to the tensile strains, shown in Fig.~\ref{strain stress}(b-f).
Most structure parameters show similar trend of variation (linear and monotonic) with respect to the tensions, especially within small strains.
However, it is noted that the out-of-plane bond angle $\theta_1$ shows non-monotonic response (negative to positive slop ratio) with respect to the biaxial tension.
This unique mechanical response indicates that there is competition between in-plane and out-of-plane atomic relaxation under tensile strain.
The plot of deformation charge density gives a more vivid understanding of the variations of structure versus the tensile strains, as shown in Fig.~\ref{charge density}.
In the analysis of Poisson's ratio, we will give further explanation about the competition mechanism of this unique mechanical response.

\begin{figure}[t]
\includegraphics[width=2.0in]{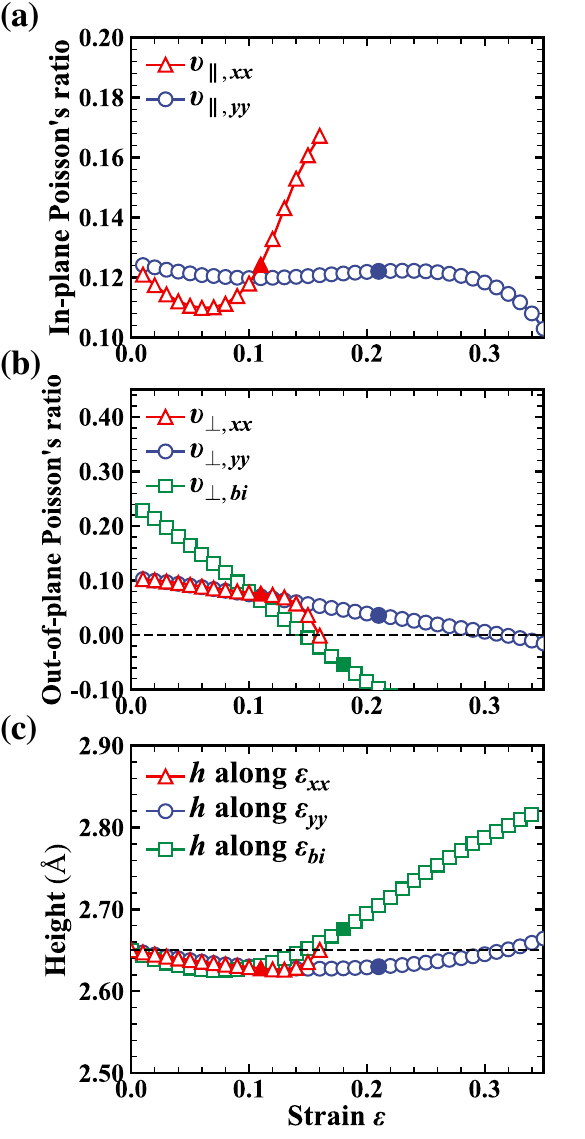}
\caption{
\label{Poisson's ratio}
Calculated (a) in-plane and (b) out-of-plane Poisson's ratio as functions of tensile strain. (c) The layer height relaxations are shown as functions of tensile strain.
The solid triangles, circles and squares indicate the condition where the peak stress could be attained along armchair, zigzag and biaxial tensions, respectively.
} 
\end{figure}

\begin{figure}[b]
\includegraphics[width=\columnwidth]{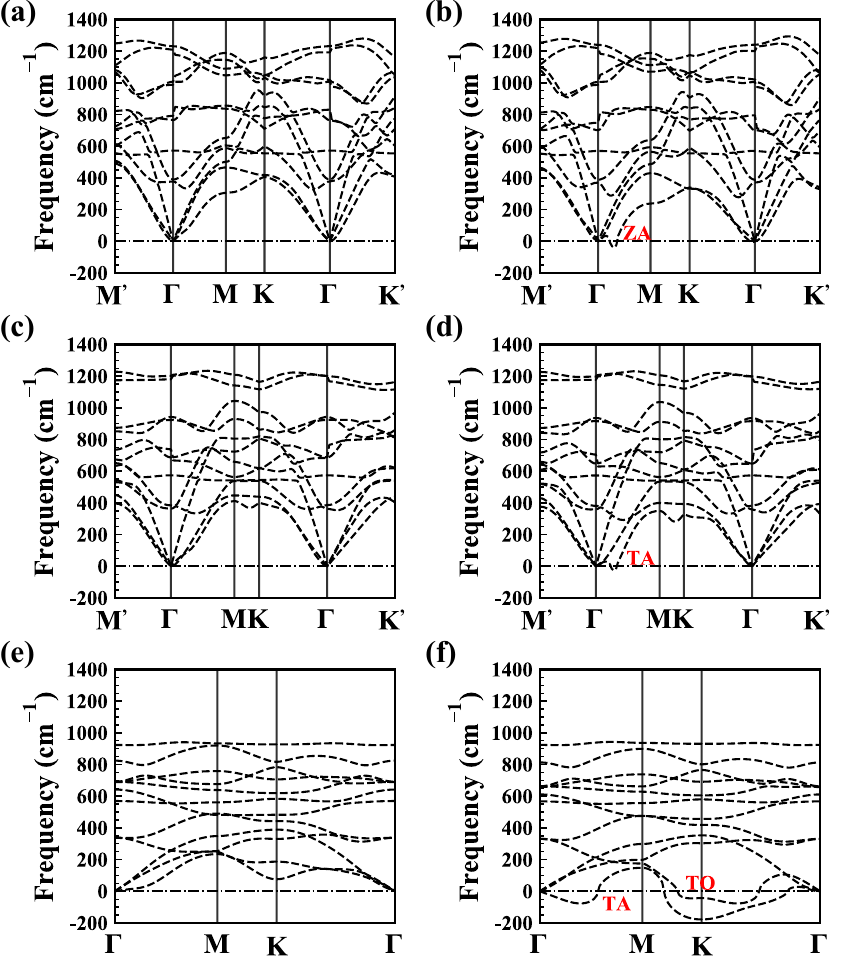}
\caption{
\label{phonon}
Calculated vibration frequency versus $\bf{k}$\ along the symmetry points in the first BZ under 
uniaxial tension (a) $\epsilon_{xx}=0.12$ and (b) $\epsilon_{xx}=0.13$;
uniaxial tension (c) $\epsilon_{yy}=0.17$ and (d) $\epsilon_{yy}=0.18$;
and biaxial tension (e) $\epsilon_{bi}=0.16$ and (f) $\epsilon_{bi}=0.17$.
The unstable phonon mode is identified to be the out-of-plane transverse phonon branch ZA for (b) armchair tension and in-plane transverse phonon branch TA for (d) zigzag tension.
} 
\end{figure}

The finite-deformation Poisson's ratio is defined as the ratio of the transverse strain to the applied strain \cite{RN160,RN925}.
The in-plane Poisson's ratio is thus well defined as $\upsilon_{\parallel, xx}=-\epsilon_{yy}^\prime/\epsilon_{xx}$ and $\upsilon_{\parallel, yy}=-\epsilon_{xx}^\prime/\epsilon_{yy}$ where the prime symbol denotes the relaxed strain. 
Nevertheless, for the out-of-plane Poisson's ratio, it depends on the determination of the deformation in the out-of-plane direction, \ti, $\epsilon_{zz}^\prime$.  
In our calculations, we make it equal the variation of layer height, namely, $\epsilon_{zz}^\prime=(h^\prime - h)/h$. 
Hence we obtain the out-of-plane Poisson's ratio $\upsilon_{\bot, xx}=-\epsilon_{zz}^\prime/\epsilon_{xx}$, $\upsilon_{\bot, yy}=-\epsilon_{zz}^\prime/\epsilon_{yy}$, and $\upsilon_{\bot, bi}=-\epsilon_{zz}^\prime/\epsilon_{bi}$. 
In Fig.~\ref{Poisson's ratio}(a), the in-plane Poisson's ratio $\upsilon_{\parallel, xx}$ and $\upsilon_{\parallel, yy}$ exhibit quantitatively different value in response to the applied strain.
The variation of $\upsilon_{\parallel, xx}$ to the strain is much intense than that of the $\upsilon_{\parallel, yy}$.
Despite the anisotropy of in-plane Poisson's ratio, 
it shows similar trend for $\upsilon_{\parallel, xx}$ and $\upsilon_{\parallel, yy}$ under applied strain (non-monotonic, negative to positive slop ratio).
In contrast, the in-plane Poisson's ratio of monolayer graphene is a monotonically decreasing function of the tensile strain \cite{RN160}.
In Fig.~\ref{Poisson's ratio}(b), a negative Poisson's ratio $\upsilon_{\bot, bi}$ in the out-of-plane direction exists under biaxial tension at $\epsilon_{bi}=0.15$. 
In order to manifest the existence of this negative out-of-plane Poisson's ratio, 
Fig.~\ref{Poisson's ratio}(c) depicts the relaxation of layer height $h$ for same set of tensile strain.
The layer height $h$ has the trend from decreasing to increasing with respect to the increase of biaxial tension.
We attribute this negative out-of-plane Poisson's ratio to the additional degrees of out-of-plane atomic relaxation introduced by the formation of interlayer bonding of CN nanosheet. 
Unlike monolayer graphene, which is a truly 2D material with single atomic layer, 
the CN nanosheet is composed of four atomic layers N-C-C-N with covalent interlayer bonding, or two buckling C-N layers.
As a consequence, the out-of-plane atomic relaxation, which is absent in monolayer graphene, will be involved in the mechanical response of CN nanosheet.
The in-plane interactions thus competing with the out-of-plane covalent bonding make a complex atomic relaxation process under tensile strain. 
We note that the different definition on the Poisson's ratio are proposed in some other reports. 
The Poisson's ratio is defined as the partial derivative of transverse strain to axial strain, namely, $\upsilon = -\partial\epsilon_{trans}/\partial\epsilon_{axial}$, in Refs.~\cite{RN948,RN947}.
With this definition, there is still a negative Poisson's ratio $\upsilon_{\bot, bi}$ under biaxial tension beyond $\epsilon_{bi}=0.08$ in the CN nanosheet. 

\begin{figure}[t]
\includegraphics[width=\columnwidth]{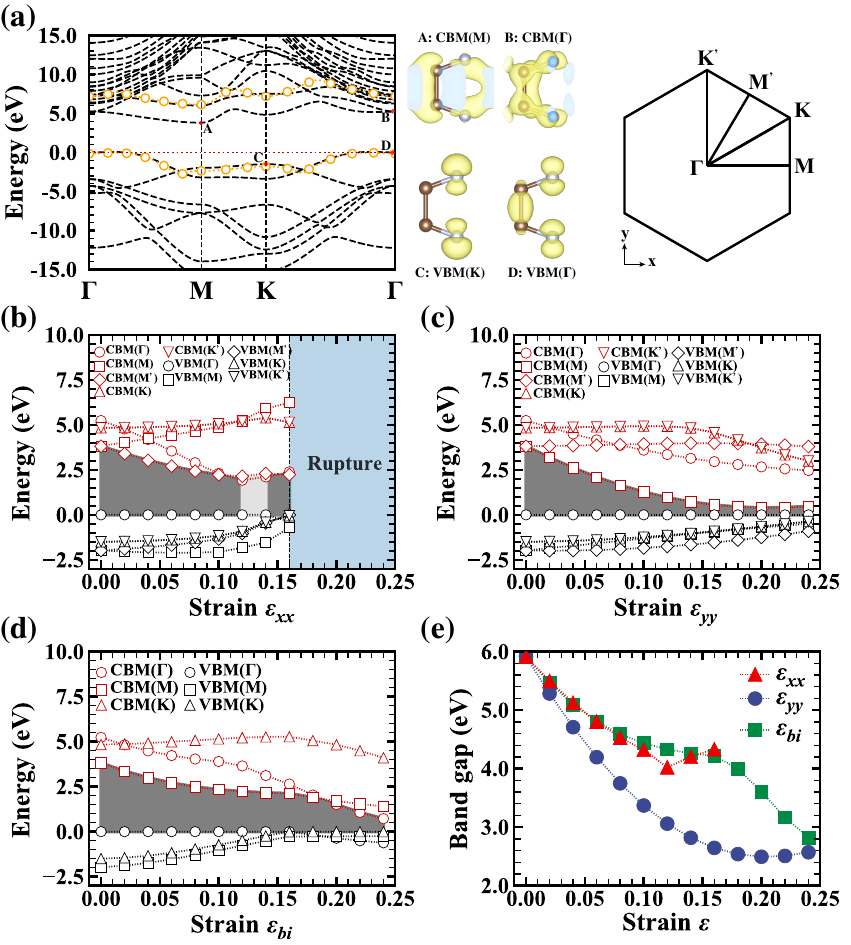}
\caption{
\label{bands}
(a) Calculated electronic band structure of CN nanosheet with PBE. 
Corrections on band gaps using $G_0W_0$ are indicated by yellow circles.
The band decomposed isosurface charge densities for lowest two conduction bands and highest two valence bands are schematically described.
The yellow and blue color represent the charge density and its cross section with periodic boundary, respectively.
The deformed first BZ under uniaxial tension is also given. 
The variations of VBM and CBM obtained with PBE under tensile strain (b) $\epsilon_{xx}$, (c) $\epsilon_{yy}$ and (d) $\epsilon_{bi}$ are shown.
The rupture along armchair tension is indicated.
Dark (light) gray regions correspond to indirect (direct) band gaps. 
(e) Variations of band gap with respect to the applied strain is summarized. 
Results obtained at the PBE level and then corrected using the scissors operator method based on the 2.1\,eV difference between PBE and PBE+$G_0W_0$ band gap at the $\Gamma$ point.
Noted that in panel (a) the Fermi energy is set to be zero.
In panel (b), (c) and (d), the energies are referenced to the VBM to illustrate the modification of band gap.
} 
\end{figure}

While the stress strain relation provides a rough indication of the ideal strength of CN nanosheet, it is still necessary to check whether the structure stability could be maintained before approaching the maximum stress, because the elastic instability may disrupt the homogeneous lattice structure on the strain path (the instability is not soft mode) \cite{RN935, RN886}.
In Fig.~\ref{phonon}, we present the anisotropic phonon dispersion curves for CN nanosheet under tensile strain.
At uniaxial strain $\epsilon_{xx}=0.12$, there is no indication of phonon instability, that is, all of phonon frequencies shown are positive [Fig.~\ref{phonon}(a)].
With the increasing of tension, the phonon instability does occur at $\epsilon_{xx}=0.13$ [Fig.~\ref{phonon}(b)].
The eigenvectors of the unstable phonon indicate that this phonon branch is the ZA mode (Fig. S1, supplementary material).
Noted that the critical strain $\epsilon_{xx}=0.13$ is beyond maximum strain $\epsilon_{xx}=0.11$ which means the ideal strength, $\sigma_{xx}^i=\sigma_{xx}^{m}=63$\,GPa, could be attained along armchair direction.
In comparison, the phonon dispersion curves maintain stable until the uniaxial strain $\epsilon_{yy}=0.17$, but then initially has a negative frequency on the TA phonon branch at $\epsilon_{yy}=0.18$.
The ideal strength is thus obtained ahead of the maximum strength, \ti, $\sigma_{yy}^i=87$ GPa, at $\epsilon_{yy}=0.17$.  
For biaxial tension, the TA branch and TO branch phonon instability simultaneously occur at $\epsilon_{bi}=0.17$.
It indicates that the failure mechanism under biaxial tension is the combination of the phonon instability and elastic instability.

Not only mechanical properties and atomic configuration but also the electronic properties of CN nanosheet, specifically its band gap, 
can be tailored through stretching.
Our results on the electronic band structures of CN nanosheet and its variation with respect to different tensile strains are presented in Fig.~\ref{bands}.  
As expected, CN nanosheet with covalent interlayer bonding is determined to be semiconductor with an indirect band gap at equilibrium state. 
It is the fact that all electrons are involved in the formation of covalent bonds in both in-plane and out-of-plane direction.
PBE predicts an indirect band gap of 3.8\,eV formed from the valance band maximum (VBM) at the $\Gamma$ point to the conduction band minimum (CBM) at the M point and a direct band gap of 5.2\,eV at the $\Gamma$ point.
With $G_0W_0$ corrections, the bands are shifted significantly in energy and the shift is not constant for different bands, resulting in a large increase of band gap at the M point than the $\Gamma$ point. 
The band gap corrections at high symmetry points are calculated to be 2.1\,eV at $\Gamma$ point, and 2.6\,eV at M and K point.
This yields 6.2\,eV and 7.3\,eV for the indirect and direct transition, respectively.
In contrast, regardless of the instability in the out-of-plane direction, 
 the monolayer structures without the interlayer bonding (Fig.~\ref{structure}(c) and Fig.~\ref{structure}(e)) are calculated to display a metallic character (Fig. S2, supplementary material). 
With this in mind, the CN nanosheet is convinced to be a semiconductor with wide indirect band gap. 

The analysis of isosurface charge density could provide a comprehensive understanding of the orbital composition of electronic band.
The highest valance state is comprised by atomic orbitals along the perpendicular direction, \ti, 
strong \so\ orbital between C atoms together with partial $p_z$ orbital of N atoms, 
and the second highest valance state is exclusively composed by partial $p_z$ orbitals perpendicular to atomic plane. 
For lowest and second lowest conduction band, both the in-plane and out-of-plane atomic orbitals take part in the band formation.
Thus, the strong interlayer bonding substantially reshape the band structure and electronic properties of CN nanosheet. 

We have plotted the variation of VBM and CBM obtained with PBE as functions of different tensile strains in Fig.~\ref{bands}(b-d). 
Generally, the band gap decreases with increasing tensile strain in the elastic region for tensile strain $\epsilon_{yy}$ and $\epsilon_{bi}$.
For armchair tension $\epsilon_{xx}$, the band gap first decreases with increasing armchair tension, passing through a minimum, 
then increases and forms a direct band gap from $\epsilon_{xx}=0.12$ to $\epsilon_{xx}=0.14$. 
This direct band gap is a consequence of the intersection of CBM between $\Gamma$ and $\mathrm{M^\prime}$.
With the increasing of these three tensions, 
there is tendency that the VBM at other high symmetry points (M, K and their splitting under uniaxial tension) are raised with regard to the VBM at $\Gamma$.
However, the variations of CBM are much more complicated and shows strong anisotropy to the applied strain.
Finally, variations of band gap with respect to the tensile strain are summarized in Fig.~\ref{bands}(e).
It is notable that these band gap results are obtained at the PBE level and then corrected using the scissors operator method based on the 2.1\,eV difference between the PBE and PBE+$G_0W_0$ band gap at the $\Gamma$ point.
Accordingly, the electronic band structure of CN nanosheet can be constructively controlled by tensile strain and 
the transition between direct and indirect band gaps can be realized.
Similar trends can be also found for 2D $X$C($X$=Si, Ge, As) nanosheets in our early work \cite{RN931}.
Thus strained CN nanosheet have important and potential applications in optoelectronics and energy engineering.

\section{Conclusion}

In summary, we have investigated the elastic, vibrational and electronic properties of the CN nanosheet in the puckered honeycomb structure with covalent interlayer bonding using first-principles simulations. 
The strong interlayer \so\ bonding between C atoms together with the hybrid intralayer $sp^3$ bonding of C-N atoms maintain the structure stability of CN nanosheet. 
It is found that CN nanosheet has an effective Young's modulus of 1013\,GPa which indicates comparable in-plane stiffness to monolayer graphene.
A negative Poisson's ratio in the out-of-plane direction under biaxial deformation, which is attributed to additional degree of atomic relaxation introduced by the interlayer bonding, is revealed. 
The highly anisotropic mechanical and electronic response of CN nanosheet to the tensile strain have been explored.
It is identified that the tensile strength of CN nanosheets are dictated by the instability of out-of-plane transverse phonon branch ZA for armchair tension $\epsilon_{xx}$, but in-plane transverse phonon branch TA for the zigzag tension $\epsilon_{yy}$. 
Our results show that the band structure of CN nanosheet could be effectively tailored through strain engineering.
This work demonstrates that CN nanosheet can be engineered through the controlling of covalent interlayer bonding and applied strain to enhance structure stability, increase buckling strength, and modify band gap, all indispensable characteristics for potential application in the nanoelectromechanical system.

\acknowledgments{
This work is supported by the Special Program for Applied Research on Super Computation of the NSFC-Guangdong Joint Fund (the second phase) under grant no. U1501501, and the Xiamen University Malaysia Research Fund, grant no. XMUMRF/2019-C3/IORI/0001.
}


\providecommand{\noopsort}[1]{}\providecommand{\singleletter}[1]{#1}%

\end{document}